\documentclass[prd,twocolumn,floats,floatfix,nofootinbib,showpacs]{revtex4}
\usepackage{graphicx}
\usepackage{dcolumn}
\usepackage{amssymb}
\usepackage{bm}
\usepackage{amsmath}
\usepackage{color}
\def\spose#1{\hbox to 0pt{#1\hss}}

\def\lta{\mathrel{\spose{\lower 3pt\hbox{$\mathchar"218$}}
\raise 2.0pt\hbox{$\mathchar"13C$}}}
\def\gta{\mathrel{\spose{\lower 3pt\hbox{$\mathchar"218$}}
\raise 2.0pt\hbox{$\mathchar"13E$}}}
\newcommand{\be}{\begin{equation}}
\newcommand{\en}{\end{equation}}
\newcommand{\bea}{\begin{eqnarray}}
\newcommand{\ena}{\end{eqnarray}}
\newcommand{\ex}{\mbox{e}}
\newcommand{\dd}{\mbox{d}}

\begin{document}
\title{Scalar Perturbations in Scalar Field Quantum Cosmology}

\author{F. T. Falciano}
\email{ftovar@cbpf.br}

\author{N. Pinto-Neto}
\email{nelsonpn@cbpf.br}

\affiliation{ICRA - Centro Brasileiro de
Pesquisas F\'{\i}sicas -- CBPF, \\ Rua Xavier Sigaud, 150, Urca, 22290-180, Rio de Janeiro, Brazil}

\date{\today}
\begin{abstract}

In this paper it is shown how to obtain the simplest 
equations for the Mukhanov-Sasaki variables describing 
quantum linear scalar perturbations in the case of scalar
fields without potential term. This was done through the
implementation of canonical transformations at the classical level, 
and unitary transformations at the quantum level, without 
ever using any classical background equation,
and it completes the simplification initiated in investigations by Langlois 
\cite{langlois}, and Pinho and Pinto-Neto 
\cite{emanuel2} for this case. These equations were then used to calculate 
the spectrum index $n_s$
of quantum scalar perturbations of a non-singular inflationary quantum background model, which starts at infinity past from flat space-time with Planckian size spacelike hypersurfaces, and inflates due to a quantum cosmological effect, until it makes an analytical graceful exit from this inflationary epoch to a decelerated classical stiff matter expansion phase.
The result is $n_s=3$, incompatible with
observations.\\
\end{abstract}


\maketitle

\section{Introduction}

The usual
theory of cosmological perturbations, with their simple
equations Ref.~\cite{mfb}, relies essentially on the assumptions that the
background is described by pure classical General Relativity (GR),
while the perturbations thereof stem from quantum fluctuations. It is
a semiclassical approach, where the background is classical and the
perturbations are quantized, and the fact that the background satisfies 
Einstein's equations is heavily used in the simplification of the equations. 
In Refs.~\cite{emanuel1,emanuel2,emanuel3}, 
which assume the validity of the Einstein-Hilbert action,
it was shown that such simple equations for quantum linear cosmological perturbations can also
be obtained without ever using any equations for the background. 
This can be accomplished through a series of canonical transformations
and redefinitions of the lapse function.
These results open the way to also quantise the background, and use these simple
equations to evaluate the evolution of the quantum linear perturbations on it.
Indeed, such results were applied to quantum bouncing backgrounds, and spectral indices 
for tensor and scalar perturbations were calculated in Refs.~\cite{ppp1,ppp2}.

The matter content used in these papers were assumed to be either a single perfect fluid or a 
single scalar field. In the case of perfect fluids, the equations were simplified
up to their simplest possible form, both for tensor and scalar perturbations.
For the case of scalar fields, this simplest form was achieved for tensor perturbations
but not for scalar perturbations. One ended in a intermediate stage that needed
further simplifications in order to be applied to quantum backgrounds Refs.~\cite{emanuel2,langlois}.

Meanwhile, a non-singular inflationary model was found Ref.~\cite{fns} containing a single
scalar field without potential term, which starts at infinity past from flat space-time with Planckian size spacelike hypersurfaces, and inflates, due to a quantum cosmological effect, until it makes an analytical graceful exit from this inflationary epoch to a decelerated classical stiff matter expansion phase. It should be interesting to investigate
if this model could generate an almost scale invariant spectrum of scalar perturbations,
as observed Ref.~\cite{wmap5}. However, without simple equations governing the evolution of the
perturbations, the investigation becomes rather cumbersome.

The aim of this paper is twofold: complete the simplification initiated in Refs.~\cite{emanuel2,langlois}, and apply it to the background described in Ref.~\cite{fns}.
In fact, after performing some canonical transformations at the classical level, 
and unitary transformations at the quantum level, we were able to obtain the simple 
equations for linear scalar perturbations of Ref.~\cite{mfb} for the case of scalar
fields without potential, without ever using any classical background equation. 
These perturbation equations were then used to calculate the spectrum index $n_s$
of the background model of Ref.~\cite{fns} yielding $n_s=3$, incompatible with
observations \cite{wmap5} ($n_s\approx 1$). Hence, even though the quantum background model has some attractive features, the model should be discarded.

The paper is organized as follows: in the next section, we briefly summarize the results of Ref.~\cite{fns}. In section III, the simplification of the second
order hamiltonian for the scalar perturbations is implemented, and the full quantization of the
system, background and perturbations, is performed. The quantum background trajectories are
then used to induce a time evolution for the Heisenberg operators describing the perturbations,
yielding simple dynamical equations for the quantum perturbations. In Section IV, we calculate the 
spectral index of scalar perturbations in the background presented in Section II, 
using the equations obtained in Section III. Section V presents our conclusions.
 
\section{Bohm-de Broglie interpretation of a quantum non-singular inflationary background model}

In this section, we first briefly highlight the main characteristics of the Bohm-de Broglie quantisation scheme, restricting our discussion to the homogeneous minisuperspace models which have a finite number of degrees of freedom. We then apply it to the quantisation of the background 
geometry with a massless scalar field without potential term. 

The Wheeler-DeWitt equation of a minisuperspace model is obtained through the Dirac quantization procedure, where the wave function must be annihilated by the operator version of the Hamiltonian constraint
\begin{equation} 
\label{bsc0}
{\cal H}({\hat{p}}^{\mu}, {\hat{q}}_{\mu}) \Psi (q) = 0 \quad.
\end{equation}
The quantities ${\hat{p}}^{\mu}, {\hat{q}}_{\mu}$ are the phase space operators
related to the homogeneous degrees of freedom of the model.
Usually this equation can be written as

\begin{equation}
\label{bsc}
-\frac{1}{2}f_{\rho\sigma}(q_{\mu})\frac{\partial \Psi (q)}{\partial q_{\rho}\partial q_{\sigma}}
+ U(q_{\mu})\Psi (q) = 0 \quad,
\end{equation}
where $f_{\rho\sigma}(q_{\mu})$ is the minisuperspace DeWitt metric of the model, whose inverse is
denoted by $f^{\rho\sigma}(q_{\mu})$.

Writing $\Psi$ in polar form, $\Psi = R \exp (iS)$, and substituting it into (\ref{bsc}),
we obtain the following equations:

\begin{equation}
\label{hoqg}
\frac{1}{2}f_{\rho\sigma}(q_{\mu})\frac{\partial S}{\partial q_{\rho}}
\frac{\partial S}{\partial q_{\sigma}}+ U(q_{\mu}) + Q(q_{\mu}) = 0 \quad,
\end{equation}
\begin{equation}
\label{hoqg2}
f_{\rho\sigma}(q_{\mu})\frac{\partial}{\partial q_{\rho}}
\biggl(R^2\frac{\partial S}{\partial q_{\sigma}}\biggr) = 0 \quad,
\end{equation}
where 

\begin{equation}
\label{hqgqp}
Q(q_{\mu}) \equiv -\frac{1}{2R} f_{\rho\sigma}\frac{\partial ^2 R}
{\partial q_{\rho} \partial q_{\sigma}} 
\end{equation}
is called the quantum potential.

The Bohm -de Broglie interpretation applied to quantum cosmology states that the trajectories $q_{\mu}(t)$ are real, independently of any observations. Equation (\ref{hoqg}) represents their Hamilton-Jacobi equation, which is the classical one  added with a quantum potential term Eq.~(\ref{hqgqp}) responsible for the quantum effects. This suggests to define

\begin{equation}
\label{h}
p^{\rho} = \frac{\partial S}{\partial q_{\rho}} ,
\end{equation}
where the momenta are related to the velocities in the usual way:

\begin{equation}
\label{h2}
p^{\rho} = f^{\rho\sigma}\frac{1}{N}\frac{\partial q_{\sigma}}{\partial t} .
\end{equation}

To obtain the quantum trajectories we have to solve the following
system of first order differential equations, called the guidance relations:
\begin{equation}\label{h3}
\frac{\partial S(q_{\rho})}{\partial q_{\rho}} =
f^{\rho\sigma}\frac{1}{N}\dot{q}_{\sigma} \quad .
\end{equation} 

Eqs. (\ref{h3}) are invariant under time reparametrization. Hence, even at the quantum level, different choices of $N(t)$ yield the same space-time geometry for a given non-classical solution $q_{\alpha}(t)$. There is no problem of time in the Bohm-de Broglie interpretation of minisuperspace quantum cosmology Ref. \cite{bola27}. We will return to this point in the next section. 

We now apply this interpretation to the situation where ${\cal H}$ in Eq. (\ref{bsc0}) is given by
\begin{equation}
H_0^{(0)}= \frac{\sqrt{2V}}{2 \ell_{Pl}e^{3\alpha}}\left(-P_{\alpha}^{2} +P_{\varphi}^2 \right) \quad ,
\end{equation}
which was worked out in Ref.~\cite{fns}. The variables are dimensionless with $\varphi$ describing the scalar field degree of freedom and $\alpha$ associated to the scale factor through $\alpha \equiv \log (a)$.
The main feature of this model is the possibility to obtain a non-singular inflationary model similar to the pre-big bang model Refs.~\cite{pbb}-\cite{veneziano}, with a minimum volume spatial section in the infinity past, or the emergent model Ref.~\cite{ellis} for flat spatial sections, without any graceful exit problem.

We take as solution of the background Wheeler-DeWitt equation, $\hat{H}_0^{(0)}\Psi(a,\varphi)=0$, a gaussian superposition of WKB solutions. The resulting wave function is (see Ref.~\cite{fns}
for details)
\begin{eqnarray}
\label{felipe}
\Psi(\alpha,\varphi) &=& 2\sqrt{\pi |h|} \biggl[\exp i\biggl(-\frac{h}{2}\left(\alpha+\varphi\right)^2 + d\left(\alpha+\varphi\right) 
+ \frac{\pi}{4}\biggr)\nonumber \\ 
&&+ \exp i\biggl(-\frac{h}{2}\left(\alpha-\varphi\right)^2+ d\left(\alpha-\varphi\right) + \frac{\pi}{4}\biggr)\biggr] \;, \qquad 
\end{eqnarray}
where $h$ and $d$ are two positive free parameters associated to the variance and the displacement of the gaussian superposition, respectively.

The norm of the wave-function is given by $R= 4 \sqrt{\pi |h|}\cos[\varphi(h\alpha-d)]$, yielding the quantum potential, Eq.~(\ref{hqgqp}),
\begin{equation}
\label{qfelipe} 
Q=  (h \alpha - d)^2-h^2\varphi^2 \quad.
\end{equation}

The guidance relations, given by Eq.~(\ref{h3}) with the choice $N=\frac{\ell_{Pl}}{\sqrt{2V}}e^{3\alpha}$, reduce to
\begin{eqnarray}
\dot{\alpha}&=&-\frac{\partial S}{\partial \alpha}\quad , \nonumber \\
\dot{\varphi}&=&\frac{\partial S}{\partial \varphi}\quad , \label{guialphaf0}
\end{eqnarray}
yielding
\begin{eqnarray}
\dot{\alpha}&=&h\alpha-d\quad , \nonumber \\
\dot{\varphi}&=&-h\varphi\quad , \label{guiphif}
\end{eqnarray}
which can be directly integrated to give
\begin{equation} \label{solf}
a=e^{\alpha}=e^{d/h}\exp (\alpha_0 e^{ht})\;\; {\rm and} \;\; \varphi=\alpha_0 e^{-ht}\quad ,
\end{equation}
where $\alpha_0$ is an integration constant. Recall that the time parameter $t$ is related to cosmic time $\tau$ through $\tau = \int dt e^{3\alpha(t)}\Rightarrow \tau-\tau_0 = {\rm Ei}(3\alpha_0e^{ht})/h$, where Ei$(x)$ is the exponential-integral function.

\begin{figure}[h]
\includegraphics[width=9cm,height=5cm]{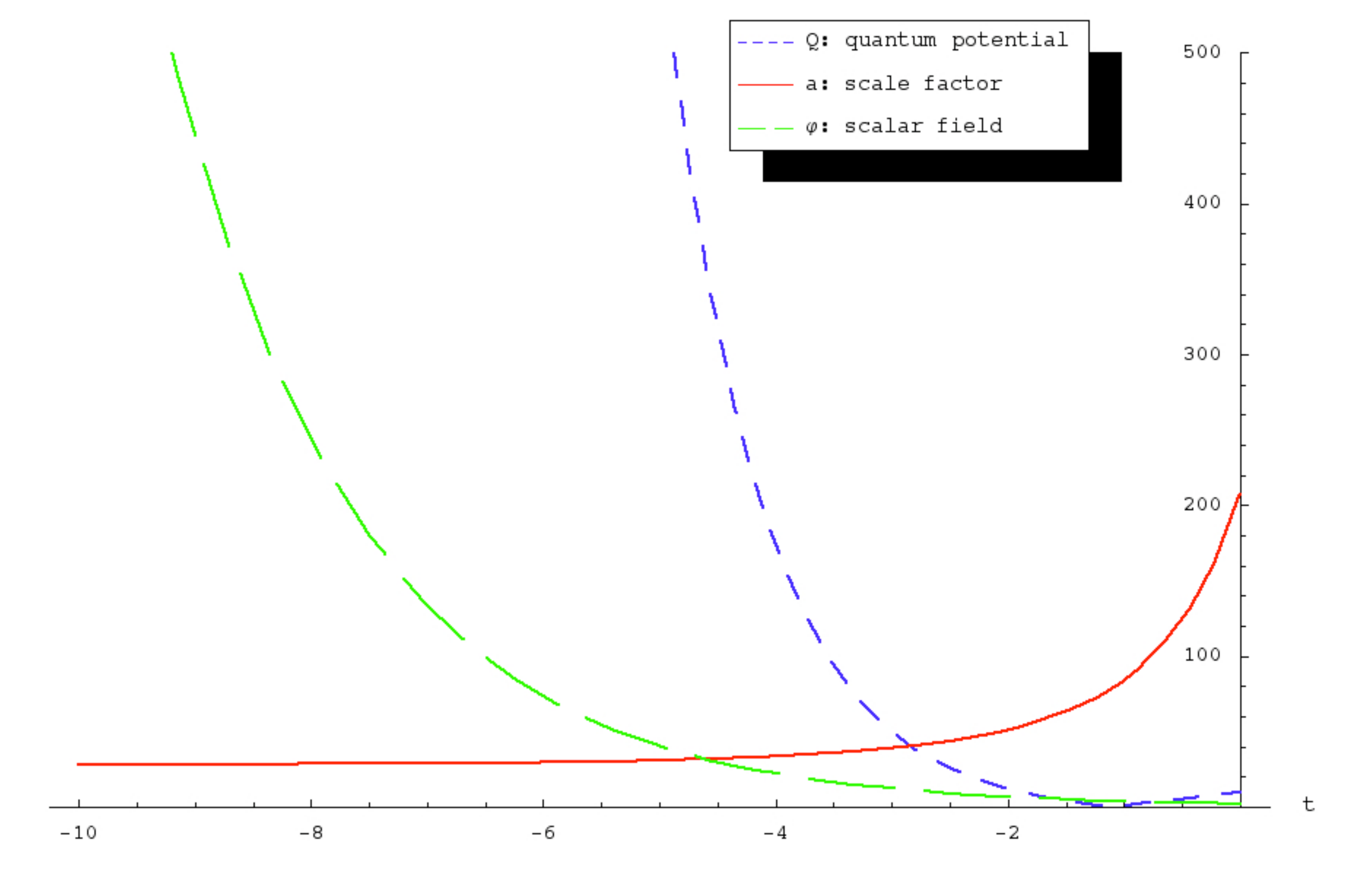}
\caption{\small{Time evolution of the background variables. The solid line describe the accelerated expansion of the scale factor from a finite minimum size $a_0= e^{d/h}$. The long-dashed line pictures the exponential decrease of the scalar field and the short-dashed line gives the decrease of the quantum potential until arriving in the classical region. The parameters were chosen to be $h=3/5$, $d=2$, and $\alpha_0=2$.}}\label{pic1}
\end{figure}

These solutions represent ever expanding non-singular models (see Figure \ref{pic1}). For $t<<0$ the Universe expands accelerately from its minimum size $a_0= e^{d/h}$ (remember that for the physical scale factor one has $a_0^{\rm phys} = \frac{\ell_{Pl}}{\sqrt{2V}} e^{d/h}$), which occurs in the infinity past $t\rightarrow-\infty$. The scalar field is very large in that phase. If $|ht|\leq\alpha_0$  is not very large, one has
\begin{widetext}
\begin{equation}
a\approx e^{\alpha_0+d/h}[1+\alpha_0 ht +(1+1/\alpha_0)(\alpha_0 h t)^2/2!+
(1+3/\alpha_0 +1/\alpha_0^2)(\alpha_0 h t)^3/3!...].
\end{equation}
Taking $\alpha_0 >> 1$, one can write $a\approx e^{\alpha_0+d/h}\exp (\alpha_0 h t)$. In that case, from $\tau = \int dt a^3(t)$, one obtains that $a\propto (\tau-\tau_0)^{1/3}$ and $\varphi \propto \ln{(\tau-\tau_0)} $, as in the classical regime. Figure 1 exhibits the bohmian trajectories and quantum potential for the parameters $h=3/5$, $d=2$, and $\alpha_0=2$. 
\end{widetext} 

\section{Simplification of the second order hamiltonian and canonical quantisation}

The conventional approach to deal with quantum cosmological perturbations is to consider a semi-classical treatment that quantise only the first order perturbations while the background 
is treated classically. Once the background dynamics has a classical evolution, one can use these equations to significantly simplify the second order lagrangian before quantising the system Ref.~\cite{mfb}. In this case, the background evolution induces a potential term that modifies the quantum dynamics of the perturbations. 

One step further is to consider quantum corrections to the background evolution itself, as in minisuperspace models, Ref.~\cite{psf}-\cite{ns}. In this case, the simplifications in the equations for the linear perturbations using the classical background cannot be implemented.  It is worth to remind that the original lagrangian is quite involved, and the use of the background equation is a key step to rewrite the system in a treatable form.
 
Recent works using technics for hamiltonian's systems Refs.~\cite{emanuel1,emanuel2,emanuel3} showed that it is also possible to simplify the full hamiltonian system by a series of canonical transformations. Their main results focus in the scalar and tensor perturbations considering the matter content of the Universe described by a perfect fluid. Even though in Ref.~\cite{langlois} and in the Appendix A of Ref.~\cite{emanuel2} it is shown a long development that significantly simplifies the hamiltonian for a scalar field with a generic potential $U(\varphi)$, there were still some delicate issues to be addressed to consistently quantise the scalar field case.

We will not reproduce the development made in these references but we will continue the development of the above mentioned Appendix. The main point to acquaint from this reference is that their simplification procedure use only canonical transformations, that guarantees the equivalence between the original and the simplified hamiltonians, independently of the background equations of motion. 

In the present work we will focus in the case of a vanishing potential $U(\varphi)$
and show how it is possible to consistently quantise simultaneously both the background and the perturbations. The background system is composed of a free massless scalar field in a spatially flat Friedmann-Lema\^itre-Robertson-Walker metric (FLRW). Since we are only interested in scalar perturbations, the perturbed metric can be written as
\begin{eqnarray}
\dd s^{2}&=&N^{2}(1+2 \phi)\dd t^{2}-NaB_{|i}\dd t \dd x^{i}+\nonumber \\
&&-a^{2}\left[(1+2\psi)\delta_{ij}-2E_{|i|j}\right]\dd x^{i}\dd x^{j} \quad .
\end{eqnarray}
The matter content is defined by a free massless scalar field $\varphi\left(t,x\right)=\varphi_0 \left(t\right) +\delta \varphi\left(t,x\right)$, where $\varphi_0$ is the background homogeneous scalar field. Using these definitions in the lagrangian density for the scalar field, namely ${\cal{L}}_{m} =\frac{1}{2}\varphi_{;\mu}\varphi^{;\mu}$, we find
\begin{eqnarray}
{\cal{L}}_m &=&\frac{\left(1-2\phi\right)}{N^2}\left(\frac{\dot{\varphi}^{2}_{0}}{2}+\dot{\varphi}\delta\dot{\varphi}\right)+\frac{\dot{\varphi}_{0}^{2}}{N^2}\left(2\phi^2-\frac{B^{|i}B_{|i}}{2}\right)+\nonumber \\
&&-\frac{\dot{\varphi}_{0}}{Na}B^{|i}\delta \varphi_{|i} +\frac{\delta \dot{\varphi}^{2}}{2N^2}-\frac{1}{2a^2}\delta \varphi^{|i}\delta\varphi_{|i} \quad .\; 
\end{eqnarray}

As our starting point, let us consider the hamiltonian (A39) of Ref.~\cite{emanuel2}
with the scalar field potential $U(\varphi)$ taken to be null,
\begin{eqnarray}\label{hh}
H&=&NH_0+\int \dd ^3x\left(-\frac{\ell^2_{Pl}P^{2}_{a}}{2a^2V}\phi+\frac{3P_{\varphi}^{2}}{a^4P_aV}\psi+\right.\nonumber\\
&&+\left.\frac{3\ell^{2}_{Pl}P_{\varphi}}{2a^4V}v \right)\tilde{\phi}_6+\Lambda_NP_N+\int \dd ^3x\Lambda_\phi\pi_\phi \quad ,
\end{eqnarray}
where $\tilde{\phi}_6=\pi_\psi$, $P_N$ e $\pi_\phi$ are first class constrains, and $v$ is the Mukhanov-Sasaki variable. The quantity $H_0$ is defined as
\begin{eqnarray}\label{ham0}
H_0&=&-\frac{\ell^2_{Pl}P^{2}_{a}}{4aV}+\frac{P^{2}_{\varphi}}{2a^3V}+\frac{1}{2a}\int \dd^3x\left( \frac{\pi^2}{\sqrt{\gamma}}+\sqrt{\gamma}v^{,i}v_{,i} \right)+\nonumber\\
&&+\left[\frac{15\ell^{2}_{Pl}P_{\varphi}^2}{4a^5V^2} -\frac{\ell^{4}_{Pl}P_{a}^{2}}{16a^3V^2}-\frac{27P_{\varphi}^{4}}{4a^7V^2P_{a}^{2}}\right]\int\dd^3x\sqrt{\gamma}v^2 \,,\qquad 
\end{eqnarray}
where $P_a$, $P_{\varphi}$ and $\pi$ are the momenta canonically 
conjugate to $a$, $\varphi_0$ and $v$, respectively, $\ell_{Pl}^{2}=\frac{8\pi G}{3}$, and $V$ is the comoving volume of the compact spatial sections, i.e. $V<\infty$. The zero order hamiltonian,
\begin{equation}
\label{h00}
H_0^{(0)}\equiv-\frac{\ell^2_{Pl}P^{2}_{a}}{4aV}+\frac{P^{2}_{\varphi}}{2a^3V}\quad ,
\end{equation}
can be used to simplify further the mass-like term for the perturbations, i.e. the function inside brackets multiplying the $v^{2}$ term. To do so, we rewrite $P_{\varphi}$ as
\[
P_{\varphi}^2=2a^3V\left(H_0^{(0)}+\frac{\ell^2_{Pl}P_a^2}{4aV}\right) \quad.
\]
Redefining the lapse function as
\[
\tilde{N}=N\left\{1+\left[\frac{15\ell^{2}_{Pl}}{2a^2V}-\frac{27}{aP_{a}^{2}}\left(H^{(0)}_0+\frac{\ell^{2}_{Pl}P_{a}^2}{2aV}\right)\right]\int\dd^3x\sqrt{\gamma}v^2\right\} \, ,
\]
and keeping only second order terms in $NH_0$, we can rewrite it as
\begin{eqnarray}\label{hh2}
NH_0&=&\tilde{N}\left[H_0^{(0)}+\frac{1}{2a}\int \dd^3x\left( \frac{\pi^2}{\sqrt{\gamma}}+\sqrt{\gamma}v^{,i}v_{,i} \right)+\right.\nonumber\\
&&\left.\frac{\ell^{4}_{Pl}P_a^2}{8a^3V^2}\int \dd ^3x\sqrt{\gamma}v^2\right]+\mathcal{O}(v^4,v^2\pi^2)\quad . \qquad 
\end{eqnarray}

Thus, by a simple redefinition of the lapse function, the mass-like term simplifies significantly. Nonetheless, it is still tricky to quantise this term due to the momentum $P_{a}$. Furthermore, the scale factor is defined on the half-line which requires additional care in specifying the Hilbert space. To deal with these two points, it is convenient to define dimensionless variables $\alpha\equiv \log{(\sqrt{2V}\ell_{Pl}^{-1}a)}$ and $\varphi\rightarrow \frac{\ell_{Pl}}{\sqrt{2}}\varphi$ which give us the following relations:
\begin{eqnarray*}
P_\alpha=-\frac{\ell_{Pl}}{\sqrt{2V}}\frac{e^{3\alpha}}{N}\dot{\alpha}&\; ,\; &\frac{\ell^{2}_{Pl}}{4V}\frac{P_a^2}{a}=\frac{\sqrt{2V}}{\ell_{Pl}}\frac{P_\alpha^2}{2e^{3\alpha}}\quad ,\\
\frac{P^{2}_{\varphi}}{2a^3V} \rightarrow \frac{\sqrt{2V}}{\ell_{Pl}}\frac{P_\varphi^2}{2e^{3\alpha}}&\; ,\; &H^{(0)}_0=\frac{\sqrt{2V}}{2\ell_{Pl}e^{3\alpha}}
\left( -P_\alpha^2+P_\varphi^2\right)\quad .
\end{eqnarray*}

With these new variables we find,
\begin{eqnarray*}
H_0&=&H_0^{(0)}+\frac{N\sqrt{2V}}{2 \ell_{Pl}e^{\alpha}}
\int \dd ^3x \sqrt{\gamma}\left( \frac{\pi^2}{\gamma}+v^{,i}v_{,i}+\frac{P_\alpha^2}{e^{4\alpha}}v^2\right). \quad 
\end{eqnarray*}
 
 To eliminate the momentum in the mass-like term we perform a canonical transformation generated by
 \begin{equation}\label{funcaogeratriz}
 \mathcal{F}=\mathcal{I}+\frac{P_\alpha}{2}\int \dd ^3x\sqrt{\gamma}\, \tilde{v}^2+e^{\tilde{\alpha}}\int \dd ^3x \, \pi\tilde{v} \quad ,
 \end{equation}
which implies
\begin{eqnarray*}
&&\alpha=\tilde{\alpha}+\frac{1}{2}\int \dd ^3x \sqrt{\gamma}\, \tilde{v}^2  \quad , \quad v=e^{\tilde{\alpha}}\tilde{v} \quad ,\\
&&\tilde{P}_\alpha=P_\alpha+e^{\tilde{\alpha}}\int \dd^3x \, \pi \tilde{v} \quad ,\quad  \tilde{\pi}=\sqrt{\gamma}\tilde{P}_{\alpha}\tilde{v}+e^{\tilde{\alpha}}\pi \quad ,\\
&&e^{3\alpha}=e^{3\tilde{\alpha}}\left(1+\frac{3}{2}\int \dd^3x\sqrt{\gamma}\, \tilde{v}^2\right)+\mathcal{O}\left(\tilde{v}^3\right)\quad.
 \end{eqnarray*}

Once more, redefining the lapse function as
\[
\tilde{N}=N\left[1-\frac{3}{2}\int \dd^3x \sqrt{\gamma}\tilde{v}^2\right] \quad ,
\]
and omitting the tilde in the new variables, the hamiltonian transforms into
\begin{eqnarray}\label{haq}
H&=&H_0+\int \dd ^3x\left(-\frac{2V}{\ell^2_{Pl}}\frac{P^{2}_{\alpha}}{e^{4\alpha}}\phi+\frac{3\sqrt{2V}}{\ell_{Pl}}\frac{P_{\varphi}^{2}}{e^{3\alpha}P_\alpha}\psi\, +\right. \nonumber\\
&&+\left. \frac{3\sqrt{2V}}{\ell_{Pl}}\frac{\sqrt{V}P_{\varphi}}{e^{4\alpha}}v \right)\pi_{\psi}+\Lambda_NP_N+\int \dd ^3x\Lambda_\phi\pi_\phi \qquad 
\end{eqnarray}
with,
\begin{equation}
\label{h0}
H_0=\frac{\sqrt{2V}}{2 \ell_{Pl}e^{3\alpha}}\left[-P_{\alpha}^{2} +P_{\varphi}^2+\int \dd ^3x \left( \frac{\pi^2}{\sqrt{\gamma}}+\sqrt{\gamma}e^{4\alpha}v^{,i}v_{,i}\right)\right] \; .
\end{equation}

The system described by this hamiltonian can be immediately quantised.
The Dirac's quantisation procedure for constrained hamiltonian systems requires that the first class constraints must annihilate the wave-function
\begin{eqnarray*}
&&\frac{\partial}{\partial N}\Psi\left(\alpha,\varphi,v,N,\phi,\psi\right)=0 \qquad ,\\
&&\frac{\delta}{\delta \psi}\Psi\left(\alpha,\varphi,v,N,\phi,\psi\right)=0 \qquad ,\\
&&\frac{\delta}{\delta \phi}\Psi\left(\alpha,\varphi,v,N,\phi,\psi\right)=0 \qquad.
\end{eqnarray*}

Thus, the wave-function must be independent of $N,\,\phi$ and $\psi$, i.e. $\Psi=\Psi\left(\alpha,\varphi,v\right)$ where  $v$ encode the perturbed degrees of freedom. Note that, due to the transformation (\ref{funcaogeratriz}),
$v$ is now the Mukhanov-Sasaki variable divided by $a$. The remaining equation is
\begin{equation}\label{vinculoh0}
\hat{H}_0\Psi\left(\alpha,\varphi,v\right)=0 \quad ,
\end{equation}
which has only quadratic terms in the momenta. 

A well known feature of the quantization of time reparametrization invariant theories is that
the state is not explicitly time dependent, hence one should find among intrinsic degrees of
freedom a variable that can play the role of time. In the perfect fluid case, the Wheeler- DeWitt's equation assumes a Schr\"odinger-like form, due to a linear term in the momenta connected with the fluid degree
of freedom. However, the hamiltonian (\ref{h0}) does not possess such linear term, rendering ambiguous
the choice of an intrinsic time variable. Notwithstanding, we still can define an evolutionary time for the perturbations if we use the Bohm-de Broglie interpretation. The procedure is similar to what is done in a semiclassical approach, where a time evolution for the quantum perturbations 
is induced from the classical background trajectory (see, e.g., Ref.~\cite{semi} for details).
Let us summarize it in the following paragraphs.

First of all, take the hamiltonian  
$NH_0$, 
with $H_0$ given in Eq. (\ref{h0}) satisfying the hamiltonian constraint
$H_0\approx 0$, and let
us solve it classically using the Hamilton-Jacobi theory. The respective Hamilton-Jacobi equation
reads
\begin{eqnarray}
&&-\frac{1}{2}\left(\frac{\partial S_T}{\partial \alpha}\right)^2+\frac{1}{2}\left(\frac{\partial S_T}{\partial \varphi}\right)^2 \nonumber \\
&&+ \frac{1}{2}\int \dd ^3x \left[ \frac{1}{\sqrt{\gamma}}
\left(\frac{\delta S_T}{\delta v}\right)^2+\sqrt{\gamma}e^{4\alpha}v^{,i}v_{,i}\right]
\; ,\quad \label{hjclass}
\end{eqnarray}
where the classical trajectories can be obtained from a solution $S_T$ of Eq.~(\ref{hjclass}) through
\begin{eqnarray}
\dot{\alpha}&=&- P_{\alpha}=-\frac{\partial S_T}{\partial \alpha}\quad , \nonumber \\
\dot{\varphi}&=& P_{\varphi}=\frac{\partial S_T}{\partial \varphi}\quad , \nonumber \\
\dot{v}&=&\frac{1}{\sqrt{\gamma}}\pi =\frac{1}{\sqrt{\gamma}}\frac{\delta S_T}{\delta v}
\quad , \label{guiclass}
\end{eqnarray}
where we have chosen $N=l_{Pl}e^{3\alpha}/\sqrt{2V}$, and hence a time parameter $t$ (a dot means derivative
with respect to this parameter), related to conformal time
through $dt \propto a^2 d\eta$.

We will now use the fact that the $v$ variable is a small perturbation over the background
variables $\alpha$ and $\varphi$, and that its back-reaction in the dynamics of the background
is negligible. In this case, one can write $S_T(\alpha,\varphi,v)$ as 
\begin{equation}
S_T(\alpha,\varphi,v)=S_0(\alpha,\varphi)+S_2(\alpha,\varphi,v) ,
\label{split}
\end{equation}
where it is assumed 
that $S_2(\alpha,\varphi,v)$ 
cannot be splitted again into a sum involving a function of the background variables alone
(which would just impose a redefinition of $S_0$). Noting that, in order to be a solution
of the Hamilton-Jacobi equation (\ref{hjclass}), $S_2$ must be at least a second order functional
of $v$ (see Ref.~\cite{hat}), then $S_2<<S_0$ as well as their partial derivatives with respect
to the background variables. Hence one obtains for the background that 

\begin{eqnarray}
\dot{\alpha}&\approx &-\frac{\partial S_0}{\partial \alpha}\quad , \nonumber \\
\dot{\varphi}&\approx &\frac{\partial S_0}{\partial \varphi} \quad . 
\label{guizeroth}
\end{eqnarray}

Inserting the splitting given in equation (\ref{split}) into equation (\ref{hjclass}), one
obtains, order by order:
\begin{widetext}
\begin{equation}
-\frac{1}{2}\left(\frac{\partial S_0}{\partial \alpha}\right)^2+\frac{1}{2}\left(\frac{\partial S_0}{\partial \varphi}\right)^2 = 0 ,
\label{split0}
\end{equation}

\begin{equation}
-\left(\frac{\partial S_0}{\partial \alpha}\right)\left(\frac{\partial S_2}{\partial \alpha}\right)+\left(\frac{\partial S_0}{\partial \varphi}\right)\left(\frac{\partial S_2}{\partial \varphi}\right)+ \frac{1}{2}\int \dd ^3x \left[ \frac{1}{\sqrt{\gamma}}
\left(\frac{\delta S_2}{\delta v}\right)^2+\sqrt{\gamma}e^{4\alpha}v^{,i}v_{,i}\right]=0,
\label{split2}
\end{equation}

\begin{equation}
-\frac{1}{2}\left(\frac{\partial S_2}{\partial \alpha}\right)^2+\frac{1}{2}\left(\frac{\partial S_2}{\partial \varphi}\right)^2 + O(4) = 0 .
\label{split4}
\end{equation}
\end{widetext}
In Eq. (\ref{split4}), the symbol $O(4)$ represents terms coming from high order corrections
to the hamiltonian (\ref{h0}). As we are interested only on linear perturbations,
this equation will not be relevant. The first equation (\ref{split0}) is the Hamilton-Jacobi 
equation of the background which solution yields, together with Eqs. (\ref{guizeroth}), the background
classical trajectories. Once one obtains the classical trajectories $\alpha(t), \varphi(t)$,
the functional $S_2(\alpha,\varphi,v)$ becomes a functional of $v$ and a function of $t$,
$S_2(\alpha,\varphi,v)\rightarrow S_2(\alpha(t),\varphi(t),v)={\bar{S}}_2(t,v)$. 
Hence equation (\ref{split2}), using Eqs. (\ref{guizeroth}), can be written as

\begin{equation}
\frac{\partial S_2}{\partial t} + \frac{1}{2}\int \dd ^3x \left( \frac{1}{\sqrt{\gamma}}
\left(\frac{\delta S_2}{\delta v}\right)^2+\sqrt{\gamma}e^{4\alpha(t)}v^{,i}v_{,i}\right)=0.
\label{split2t}
\end{equation}

Equation (\ref{split2t}) can now be understood as the Hamilton-Jacobi equation coming from the
hamiltonian
\begin{equation}
\label{htau}
H_2=\frac{1}{2}\int \dd ^3x \left( \frac{\pi^2}{\sqrt{\gamma}}+\sqrt{\gamma}e^{4\alpha(t)}v^{,i}v_{,i}\right) ,
\end{equation}
which is the generator of time $t$ translations (and not anymore constrained to be null).

If one wants to quantize the perturbations, the correspoding Schr\"odinger equation should be
\begin{equation}
i\frac{\partial \chi}{\partial t}=\hat{H}_2\chi\, ,
\end{equation}
where $\chi$ is a wave functional depending on $v$ and $t$, and the dependences
of $\hat{H}_2$ on the background variables are understood as a dependence on $t$.

Let us now go one step further and quantize both the background and perturbations.
When the background is also quantised, this procedure can also be implemented
in the framework of the Bohm-de Broglie interpretation of quantum theory, where there is a definite notion of trajectories as well, the bohmian trajectories. In order to do that, we 
first note that Eqs. (\ref{vinculoh0}) and (\ref{h0}) imply that
\begin{equation}
\label{split-h0}
(\hat{H}_0^{(0)}+\hat{H}_2) \Psi = 0, 
\end{equation}
where
\begin{eqnarray}
&&\hat{H}_0^{(0)}=-\frac{\hat{P}_\alpha^2}{2}+\frac{\hat{P}_\varphi^2}{2} \quad , \\
&&\hat{H}_2=\frac{1}{2}\int \dd ^3x \left( \frac{\hat{\pi}^2}{\sqrt{\gamma}}+\sqrt{\gamma}e^{4\hat{\alpha}}\hat{v}^{,i}\hat{v}_{,i}\right) \quad .
\end{eqnarray}

We write the wave functional $\Psi$ as $\Psi=\exp(A_T+iS_T)\equiv R_T\exp(iS_T)$,
where both $A_T$ and $S_T$ are real functionals.
Inserting it in the Wheeler-DeWitt equation (\ref{split-h0}), the two real
equations we obtain are
\begin{widetext}
\begin{equation}
\label{Thoqg}
-\frac{\partial}{\partial \alpha}
\biggl(R_T^2\frac{\partial S_T}{\partial \alpha}\biggr) 
+\frac{\partial}{\partial \varphi}
\biggl(R_T^2\frac{\partial S_T}{\partial \varphi}\biggr)
+\int\frac{\dd^3 x}{\sqrt{\gamma}}\frac{\delta}{\delta v}
\biggl(R_T^2\frac{\delta S_T}{\delta v}\biggr)= 0 \quad,
\end{equation}

\begin{equation}
-\frac{1}{2}\left(\frac{\partial S_T}{\partial \alpha}\right)^2+\frac{1}{2}\left(\frac{\partial S_T}{\partial \varphi}\right)^2 
+ \frac{1}{2}\int \dd ^3x \left( \frac{1}{\sqrt{\gamma}}
\left(\frac{\delta S_T}{\delta v}\right)^2+\sqrt{\gamma}e^{4\alpha}v^{,i}v_{,i}\right)
+\frac{1}{2R_T}\left(\frac{\partial ^2 R_T}{\partial \alpha ^2}-
\frac{\partial ^2 R_T}{\partial \varphi ^2}\right) -
\frac{1}{2}\int\frac{\dd^3 x}{\sqrt{\gamma}}\frac{1}{R_T}\frac{\delta^2 R_T}{\delta v^2}= 0
\; .\quad \label{hjquant}
\end{equation}
\end{widetext}
These two equations correspond to equations (\ref{hoqg2}) and (\ref{hqgqp}), respectively.

The bohmian guidance relations are the same as in the classical case,

\begin{eqnarray}
\dot{\alpha}&=&- P_{\alpha}=-\frac{\partial S_T}{\partial \alpha}\quad , \nonumber \\
\dot{\varphi}&=& P_{\varphi}=\frac{\partial S_T}{\partial \varphi}\quad , \nonumber \\
\dot{v}&=&\frac{1}{\sqrt{\gamma}}\pi =\frac{1}{\sqrt{\gamma}}\frac{\delta S_T}{\delta v}
\quad , \label{guiq}
\end{eqnarray}
with the difference that the new $S_T$ satisfies a Hamilton-Jacobi equation different
from the classical one due to the presence of the quantum potential terms (the two
last terms in Eq. (\ref{hjquant})), which are responsible for the quantum effects.

We have again made the choice $N\propto e^{3\alpha}$. Whether this procedure is 
unambiguously independent
on the choice of the lapse function is a delicate point. Indeed, in a general framework 
(the full superspace), the bohmian evolution
of three-geometries may not even form a four-geometry (a spacetime) in the sense
described in Refs.~\cite{santini0,tese,cons,shtanov}, although the theory remains consistent
(Refs. \cite{tese,cons}), and its geometrical properties depends on the choice
of the lapse function.
However, in the case of homogeneous spacelike hypersurfaces, a preferred foliation
of spacetime is selected, the one where the time direction is perpendicular to the 
Killing vectors of these hypersurfaces. In this case, once one has chosen this
preferred foliation, one can prove that the residual ambiguity in the lapse function
(which is now independent of space coordinates)
is geometrically irrelevant for the Bohmian trajectories (see Ref.~\cite{bola27}).
This is also true when linear perturbations are present, where the hamiltonian constraints reduce
to a single one, and the super-momentum constraint can be solved, as it was shown in 
Ref.~\cite{ppp2}. Again, the lapse function is just a time function. In this case, the bohmian
quantum background trajectories can be obtained without geometrical ambiguities \cite{bola27},
and they can be used to induce a time dependence on the perturbation quantum state, as we will see.

Let us assume, as in the classical case, that we can split 
$A_T(\alpha,\varphi,v)=A_0(\alpha,\varphi)+A_2(\alpha,\varphi,v)$
implying that $R_T(\alpha,\varphi,v)=R_0(\alpha,\varphi)R_2(\alpha,\varphi,v)$, and
$S_T(\alpha,\varphi,v)=S_0(\alpha,\varphi)+S_2(\alpha,\varphi,v)$, and that $A_2 << A_0$,
$S_2 << S_0$, together with their derivatives with respect to the background variables. 
The approximate guidance relations are
\begin{eqnarray}
\dot{\alpha}&\approx &-\frac{\partial S_0}{\partial \alpha}\quad , \nonumber \\
\dot{\varphi}&\approx &\frac{\partial S_0}{\partial \varphi}\quad \; \; , 
\label{guizerothq}
\end{eqnarray}
and the zeroth order terms of Eqs. (\ref{Thoqg}) and (\ref{hjquant}) read

\begin{equation}
\label{0Thoqg2}
-\frac{\partial}{\partial \alpha}
\biggl(R_0^2\frac{\partial S_0}{\partial \alpha}\biggr) 
+\frac{\partial}{\partial \varphi}
\biggl(R_0^2\frac{\partial S_0}{\partial \varphi}\biggr)
\approx 0 \quad,
\end{equation}
\begin{equation}
-\frac{1}{2}\left(\frac{\partial S_0}{\partial \alpha}\right)^2+\frac{1}{2}\left(\frac{\partial S_0}{\partial \varphi}\right)^2 
+\frac{1}{2R_0}\left(\frac{\partial ^2 R_0}{\partial \alpha ^2}-
\frac{\partial ^2 R_0}{\partial \varphi ^2}\right) \approx 0
\; .\quad \label{0hjquant}
\end{equation}
which, again, correspond to Eqs. (\ref{hoqg2}) and (\ref{hqgqp}) for the background, respectively.

A solution $(S_0,R_0)$ of Eqs. (\ref{0Thoqg2}) and (\ref{0hjquant}) yield a bohmian quantum
trajectory for the background through Eq. (\ref{guizerothq}). If $S_0$ and $R_0$ are obtained from
Eq. (\ref{felipe}), then the bohmian trajectories will be given by Eq. (\ref{solf}).

As in the classical case, once one obtains the bohmian quantum trajectories $\alpha(t), \varphi(t)$,
the functionals $S_2(\alpha,\varphi,v)$, $A_2(\alpha,\varphi,v)$ become  functionals of $v$ and functions of $t$,
$S_2(\alpha,\varphi,v)\rightarrow S_2(\alpha(t),\varphi(t),v)={\bar{S}}_2(t,v)$,
$A_2(\alpha,\varphi,v)\rightarrow A_2(\alpha(t),\varphi(t),v)={\bar{A}}_2(t,v)$.

Defining $\chi(\alpha,\varphi,v)\equiv R_2(\alpha,\varphi,v)\exp(iS_2(\alpha,\varphi,v))$,
writing it as
\begin{equation}
\label{defF}
\chi(\alpha,\varphi,v)=\int{\dd \lambda}\,G(\lambda,v)F(\lambda,\alpha,\phi)\quad ,
\end{equation}
where $F$ satifies
\begin{equation}
\label{F}
\frac{1}{2}\left(\frac{\partial^2 F}{\partial \alpha^2}-\frac{\partial^2 F}{\partial \varphi^2}\right)+\frac{1}{R_0}\left(\frac{\partial R_0}{\partial \alpha}\frac{\partial F}{\partial \alpha}-\frac{\partial R_0}{\partial \varphi}\frac{\partial F}{\partial \varphi}\right)=0\, , 
\end{equation}
and $G$ is an arbitrary functional of $v$, which also depends on an integration constant $\lambda$,
then the next-to-leading-order terms of Eqs. (\ref{Thoqg}) and (\ref{hjquant}) read
\begin{widetext}
\begin{equation}
\label{Thoqg2}
\frac{\partial {\bar{R}}_2^2}{\partial t}
+\int\frac{\dd^3x}{\sqrt{\gamma}}\frac{\delta}{\delta v}
\biggl({\bar{R}}_2^2\frac{\delta {\bar{S}}_2}{\delta v}d^3 x\biggr)= 0 \quad,
\end{equation}

\begin{equation}
\frac{\partial {\bar{S}}_2}{\partial t}
+ \frac{1}{2}\int \dd ^3x \left( \frac{1}{\sqrt{\gamma}}
\left(\frac{\delta {\bar{S}}_2}{\delta v}\right)^2+\sqrt{\gamma}e^{4\alpha(t)}v^{,i}v_{,i}\right)
-\frac{1}{2}\int\frac{\dd^3 x}{{\bar{R}}_2\sqrt{\gamma}}\frac{\delta^2 {\bar{R}}_2}{\delta v^2}
= 0
\; ,\quad \label{hjquant2}
\end{equation}
where ${\bar{R}}_2(t,v)\equiv\exp({\bar{A}}_2(t,v))$.
\end{widetext}
 In order to obtain these equations we used that
\begin{equation}
-\left(\frac{\partial S_0}{\partial \alpha}\right)\left(\frac{\partial S_2}{\partial \alpha}\right) +
\left(\frac{\partial S_0}{\partial \varphi}\right)\left(\frac{\partial S_2}{\partial \varphi}\right)
= \frac{\partial {\bar{S}}_2}{\partial t},
\end{equation}
and the same for $R_2$ and ${\bar{R}}_2$.

These two equations can be grouped into a single Schr\"odinger equation
\begin{equation}
\label{xo}
i\frac{\partial \bar{\chi}}{\partial t}=\hat{H}_2\bar{\chi}\, ,
\end{equation}
where $\bar{\chi}(t,v)=\chi(\alpha(t),\varphi(t),v)$ is a wave functional depending on $v$ and $t$,
and, as before, the dependences
of $\hat{H}_2$ on the background variables are understood as a dependence on $t$.

For the specific example of section II, Eq. (\ref{felipe}), one possible solution of Eq.(\ref{F}) yields for $\chi$ through Eq.(\ref{defF})
\begin{widetext}
\begin{equation}
\label{Fsol}
\chi(\alpha,\varphi,v)=\frac{1}{R(\alpha,\varphi)}\int{\dd \lambda}\,G(\lambda,v)\exp\left\{ \frac{(\alpha+\varphi-d/h)^2}{2 \lambda}+\frac{\lambda h^2 (\alpha-\varphi-d/h)^2}{8}\right\}\quad .
\end{equation}
\end{widetext}
From solution (\ref{Fsol}), we can construct $\bar{\chi}(t,v)\equiv \chi(\alpha(t),\varphi(t),v)$ solution
of Eq. (\ref{xo}). Note that, as $G$ is an arbitrary functional of $v$ and the real parameter $\lambda$,
the functional $\bar{\chi}(t,v)$ constructed from (\ref{Fsol}) via 
$\bar{\chi}(t,v)\equiv \chi(\alpha(t),\varphi(t),v)$
is also an arbitrary functional of $t$ and $v$
(even though $\chi(\alpha,\varphi,v)$ in (\ref{Fsol}) is not arbitrary in $\alpha$ and $\varphi$).

During our procedure, we have supposed that the evolution of the background is independent of the perturbations. 
This no back-reaction assumption is based on the fact that
terms induced by the linear perturbations in the zeroth order hamiltonian 
are negligible, which should be the case when one assumes that quantum perturbations are initially in
a vacuum quantum state, as it is argued in Ref.~\cite{halliwell}.
We will come back to this point in the conclusion.

Once one obtains the quantum trajectories for the background variables, they can be used to define a time dependent unitary transformation for the perturbative sector. This unitary transformation takes the vector $|\chi\rangle $ into $|\xi\rangle=U|\chi\rangle$, i.e. $|\chi\rangle=U^{-1}|\xi\rangle$. With respect to this transformation the hamiltonian is taken into $\hat{H}_2\longrightarrow \hat{H}_{2U}$ with
\begin{equation}
i\frac{\dd}{\dd t}|\xi\rangle=\hat{H}_{2U}|\xi\rangle =\left(U\hat{H}_2U^{-1}-iU\frac{\dd }{\dd t}U^{-1} \right)|\xi\rangle\quad.
\end{equation}


Let us define this unitary transformation by
\begin{equation}\label{Utransf}
U=e^{iA}e^{-iB}
\end{equation}
with,
\begin{eqnarray}
&&A=\frac{1}{2}\int \dd^3x\sqrt{\gamma}\frac{\dot{a}}{a^3}\hat{v}^2 \qquad ,\\
&&B=\frac{1}{2}\int \dd^3x\left(\hat{\pi} \hat{v}+\hat{v}\hat{\pi}\right)\log(a) \qquad .
\end{eqnarray}

Remember that the time derivative, $\dot{a}=\frac{d a}{d t}$, is taken with respect to the parametric time $t$ related to the cosmic time $\tau $ by $\dd \tau=N \dd t\propto a^3\dd t$. In these expressions, the scale factor $a=a(t)$ should be understood as a function of time, instead of an operator, since we suppose that the background quantum equations have already been solved. Thus, $a=a(t)$ should be taken as the bohmian trajectory associated with equations $\hat{H}_0^{(0)}|\phi\rangle=0$.\\

Naturally, the $\hat{\pi}$ e $\hat{v}$ operators do not commute with the unitary transformation. Using the following relations
\begin{eqnarray*}
e^{iA}\, \hat{v}\, e^{-iA}=\hat{v} &\quad , \quad &  e^{iA}\, \hat{\pi}\, e^{-iA}=\hat{\pi}-\frac{\dot{a}}{a^3}\sqrt{\gamma}\, \hat{v}\\
e^{-iB}\,\hat{v}\, e^{iB}=a^{-1} \, \hat{v}& \quad , \quad &e^{-iB}\, \hat{\pi}\, e^{iB}=a\hat{\pi}  \quad .
\end{eqnarray*}
we can calculate the transformed hamiltonian as 
\begin{equation}\label{h2u}
\hat{H}_{2U}=\frac{a^2}{2}\int \dd^3x \left[ \frac{\hat{\pi}^2}{\sqrt{\gamma}}+\sqrt{\gamma}\,\hat{v}^{,i}\hat{v}_{,i}-\left(\frac{\ddot{a}}{a^5}-2\frac{\dot{a}^2}{a^6} \right)\sqrt{\gamma}\,\hat{v}^2\right] \qquad 
\end{equation}

Note that the unitary transformation $U$ takes us back to the Mukhanov-Sasaki variable. 

Recalling that $\dd t=a^{-2}\dd \eta$, where $\eta$ is the conformal time, we have $\dot{a}=a^2a'$ and $\ddot{a}=a^4a''+2a^3a'^2$, and the hamiltonian can be recast as
\begin{equation}\label{hquantica}
\hat{H}_{2U}= \frac{a^2}{2}\int \dd^3x \left[ \frac{\hat{\pi}^2}{\sqrt{\gamma}}+\sqrt{\gamma}\,\hat{v}^{,i}\hat{v}_{,i}-\frac{a''}{a}\sqrt{\gamma}\,\hat{v}^2\right]\; .
\end{equation}

So far our analysis has been made in the Schr\"odinger picture but now it is convenient to describe the dynamics using the Heisenberg representation. The equations of motion for the Heisenberg operators
are written as
\begin{eqnarray*}
&&\dot{\hat{v}}=-i\left[\hat{v},\hat{H}_{2U}\right]=a^2\frac{\hat{\pi}}{\sqrt{\gamma}} \qquad ,\\
&&\dot{\hat{\pi}}=-i\left[\hat{\pi},\hat{H}_{2U}\right]=a^2 \sqrt{\gamma}\left(\hat{v}^{,i}_{\phantom a ,i}+\frac{a''}{a}\hat{v}\right) \qquad.
\end{eqnarray*}

Combining these two equations and changing to conformal time, we find
the following equations for the operator modes of wave number $k$, $v_k$:
\begin{equation}\label{eqv}
v_k''+ \left(k^2-\frac{a''}{a}\right)v_k \quad =0 \qquad .
\end{equation}

This is the same equation of motion for the perturbations known in the literature in the absence of a scalar field potential Ref.~\cite{mfb}. The crucial point is that we have not used the background equations of motion. Thus we have shown that Eq. (\ref{eqv}) is well defined, independently of the background dynamics, and it is correct even if we consider quantum background trajectories.

Note, however, that this result was obtained using a specific subclass of wave functionals which satisfies
the extra condition Eq. (\ref{F}). What are the physical assumptions behind this choice?

When one approaches the classical limit, where $R_0$ is a slowly varying function of
$\alpha$ and $\varphi$, condition (\ref{F}) reduces to 

\begin{equation}
\label{Fc}
\frac{\partial^2 F}{\partial \alpha^2}-\frac{\partial^2 F}{\partial \varphi^2}\approx 0\, .
\end{equation}
If Eq. (\ref{Fc}) were not satisfied, one would not obtain anymore the usual Schr\"odinger equation
for quantum perturbations in a classical background (which arises when $R_0$ is a slowly varying 
function of $\alpha$ and $\varphi$), due to extra terms in Eqs. (\ref{Thoqg2}) and
(\ref{hjquant2}): there would be corrections originated 
from some quantum entanglement between the background and the perturbations, even when
the background is already classical, which would spoil
the usual semiclassical approximation. This could be a viable possibility driven by a different
type of wave functional than the one considered here, but it seems that our Universe is not so complicated.
In fact, the observation that the simple semiclassical model without this sort of
entaglement works well in the real Universe indicates something about the wave functional of 
the Universe\cite{hartle27}\footnote{In these references, it is pointed out how the features
of our Universe we take for granted (classicality, separability) impose severe restrictions
on the initial wave function of the Universe. In fact, our Universe could have been highly
nonclassical, completely entagled, even when it is large, depending on the features of
this initial wave solution.}. In other words, the validity 
of the usual semiclassical approximation imposes Eq. (\ref{Fc}).

When $R_0$ is not slowly varying and quantum effects on the background become important
causing the bounce, the two last terms of condition (\ref{F}) cannot be neglected. They would also
induce extra terms in Eqs. (\ref{Thoqg2}) and
(\ref{hjquant2}), again originated 
from some quantum entanglement between the background and the perturbations, but now in
the background quantum domain, and the
final quantum equation (\ref{eqv}) for the perturbations we obtained would not be valid around the
bounce. In this case, there is no observation indicating which class of wave functionals
one should take and our choice in this no man's land resides only on assumptions of simplicity:
there is no quantum entaglement between the background and the perturbations in the entire
history of the Universe. This is the physical hypothesis behind the choice of the specific
class os wave functionals satisfying condition (\ref{F}).

In the next section we will apply the above formalism implying Eq. (\ref{eqv})
to the specific example described in section II.\\


\section{Application of the formalism}

We will now use Eq. (\ref{eqv}) to evaluate the spectral index of scalar perturbations
in the quantum background described by Eq. (\ref{solf}). The potential $V\equiv a''/a$ reads
\begin{eqnarray}\label{potential}
V\equiv\frac{a''}{a}&=&\frac{1}{a^4}
\biggl[\frac{\ddot{a}}{a}-\biggl(\frac{\dot{a}}{a}\biggr)^2 \biggr]\nonumber\\
&=&\frac{\alpha_0 h^2\exp(ht)[1-\alpha_0\exp(ht)]}{a^4}.
\end{eqnarray}

Defining $u_k\equiv v_k/a$, Eq. (\ref{eqv}) in terms of the $t$ variable
can be written as (from now on we will omit the index $k$),
\begin{equation}\label{equ}
\ddot{u}+k^2a^4 u =0 \qquad .
\end{equation}
When $ht<<0$, we can approximate $a\approx\exp(d/h)[1+\alpha_0\exp(ht)]$, and the general solution reads 
\begin{equation}\label{solu}
u=A_+(k)J_{\nu}(z)-A_-(k)J_{-\nu}(z)\qquad ,
\end{equation}
where $J$ is the Bessel function of the first type, $\nu =i2k\exp(2d/h)/h$
and $z=4\alpha_0^{1/2}k\exp(2d/h+ht/2)/h$. At $t\rightarrow -\infty$, when
the scale factor becomes constant and spacetime is flat, one can impose
vacuum initial conditions 
\be v_\mathrm{ini} = \frac{\ex^{i
k\eta}}{\sqrt{k}}, \label{v1ini} ,
\en
which implies that $A_+(k)=0$, and $A_-(k)\propto k^{-1/2}\exp[i2k\ln(k)\exp(2d/h)/h]$.
Hence, $v$ in this region reads
\begin{equation}\label{solvI}
v_I=aA_-(k)J_{-\nu}(z)\qquad .
\end{equation}

The solution can also be expanded in powers of $k^2$ according to
the formal solution (see Ref.~\cite{mfb})
\begin{eqnarray}
\frac{v}{a} & \simeq & A_1(k)\biggl[1 - k^2 \int^t \frac{\dd\bar
  \eta}{a^2\left(\bar \eta\right)} \int^{\bar{\eta}}
  a^2\left(\bar{\bar{\eta}}\right)\dd\bar{\bar{\eta}}\biggr]\nonumber
  \\ &+& A_2(k) \biggl[\int^\eta\frac{\dd\bar{\eta}}{a^2} - k^2
  \int^\eta \frac{\dd\bar{\eta}}{a^2} \int^{\bar{\eta}} a^2
  \dd\bar{\bar{\eta}} \int^{\bar{\bar{\eta}}}
  \frac{\dd\bar{\bar{\bar{\eta}}}}{a^2} \biggr]+...,\cr & & \label{solform}
\end{eqnarray}
When the mode is deep inside the potential,
$k^2<<V$, we can neglect the $k^2$ terms yielding
\begin{equation}\label{solvII}
v_{II}\approx a\biggl[A_1(k)+A_2(k)\int^\eta\frac{\dd\bar{\eta}}{a^2}\biggr]=a\biggl[A_1(k)+A_2(k)t\biggr]\qquad .
\end{equation}

We can now perform the matching of $v_I$ with $v_{II}$ in order to calculate
$A_1(k)$ and $A_2(k)$. As we are interested on large scales, $k<<1$, this matching
can still be made when $ht<<0$. In this region one has $V\approx\alpha_0 h^2 \exp(ht-4d/h)$,
yielding the matching time
\begin{equation}\label{mattime}
ht_M = \ln\biggl(\frac{k^2\exp(4d/h)}{\alpha_0 h^2}\biggr)\qquad .
\end{equation}
Note that the potential crossing condition relating the wave number $k$ and the time $t_M$ of the crossing is logarithmic. In fact, since in this region the scale factor is almost constant, the wave number is also logarithmically related to the conformal time. This dependence is drastically different from the slow roll scenario, where the conformal time of potential crossing is inversely proportional to the wave number, $k \propto 1/ \eta_{M}$.

Performing the matching at this time and taking the leading order term in $k$,
one obtains that
\begin{equation}\label{a1a2}
A_1(k)=k^{-1}A_2(k)\propto k^{-1/2}\exp[i6k\ln(k)\exp(2d/h)/h].
\end{equation}

\begin{figure}[h]
\includegraphics[width=8cm,height=4cm]{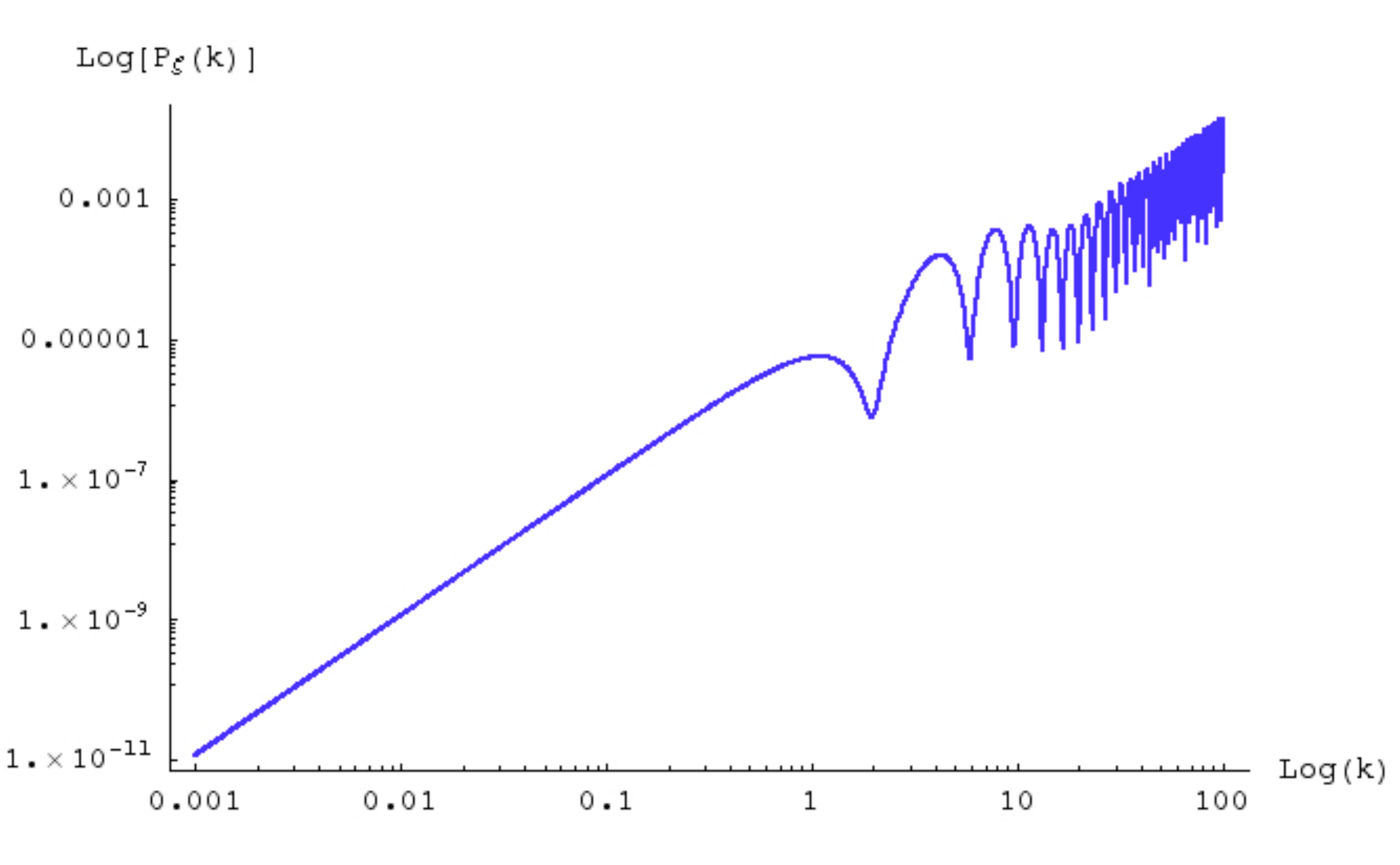}
\caption{The power spectrum $\mathcal{P}_\Phi$ calculated numerically. The numerical integration was carried out with $h=d= 3\times 10^{2}$ and $\alpha_0= 1$. Since this is a log-log plot, one can immediately check that $\mathcal{P}_\Phi \propto k^{2}$ for small $k$.}\label{fig2}
\end{figure}

Note that solution (\ref{solform}) is valid everywhere, hence we can use it 
in the period when the scale factor evolution becomes classical. During this period,
unless for some fine tuning, the mode is also deep inside the potential and one can
use Eq. (\ref{solvII}) to calculate the Bardeen potential $\Phi$ through the classical
equation Ref.~\cite{mfb}
\begin{equation}\label{bardeen}
\Phi=-\frac{(\epsilon+p)^{1/2}z}{k^2}\biggl(\frac{v}{z}\biggr)',
\end{equation}
where $z\equiv a^2(\epsilon+p)^{1/2}/{\cal{H}}$. For the case of a scalar
field without potential (stiff matter), $z\propto a$, yielding
\begin{equation}\label{bardeen2}
\Phi\propto A_1(k)+\frac{A_2(k)}{k^2 a^4},
\end{equation}
one constant and one decaying mode, as usual. The transition to radiation dominated
and matter dominated phases may alter the amplitudes but not the spectrum.
The power spectrum
\begin{equation}
\mathcal{P}_\Phi \equiv \frac{2 k^3}{\pi^2}
\left| \Phi \right|^2 \propto k^{n_{_\mathrm{S}}-1},
\label{powspec}
\end{equation}
yields for the spectral index, from the value of $A_1(k)$ in the constant mode given in Eq. (\ref{a1a2}), the value $n_s=3$, contrary to observational results Ref.~\cite{wmap5}. This power law dependence was checked numerically as can be seen by figure \ref{fig2}. Hence, the model cannot describe the primordial era of our Universe.

\section{Conclusion}\label{conclusion}

In this paper we were able to obtain the simple equation for linear scalar perturbations of Ref.~\cite{mfb} for the case of a scalar field without potential. The simplification procedure was carried out without ever using any classical background equation. Instead, by a series of canonical transformations and redefinitions of the lapse function we are able to put the hamiltonian in a form susceptible to quantization. 

However, contrary to the perfect fluid case, the scalar field minisuperspace model has no natural way to define a time variable since its hamiltonian constraint does not contain a linear term in the momenta. Nevertheless, if one assumes there is no back-reaction, we have shown how to bypass this problem using the quantum background bohmian trajectories. The quantum background dynamics in the Bohm-de Broglie interpretation naturally provides an evolutionary time to the perturbative sector, similarly to what is done at the semiclassical level through the classical background trajectories \cite{semi}. 

These perturbation equations were then used to calculate the spectrum index $n_s$ of the background model of Ref.~\cite{fns} yielding $n_s=3$, incompatible with observations Ref.~\cite{wmap5} ($n_s\approx 1$). This result is intimately related to the logarithmically dependence of the wave number to the potential crossing time, see eq. (\ref{mattime}). As a consequence, the model should be discarded. This is an example of an inflationary model without (almost) scale invariant scalar perturbations.

The no back-reaction hypothesis we have used was justified through
the assumption that the perturbations are in a quantum vacuum state 
initially \cite{halliwell}. One could verify the consistency
of such hypothesis by checking whether the perturbations calculated under this assumption
never departs the linear regime in the region where the background is influenced by quantum
effects. This check was done in other frameworks (see Ref.~\cite{ppp2}), 
where self-consistency was 
verified. This self-consistency check, however, was not implemented here because the model 
studied in section IV does not present a scale invariant spectrum for long-wavelength
perturbations, and the model should be discarded without the need of calculating the
amplitude of perturbations.

We have also assumed that there is no quantum entanglement in such a way that the background disturbs
the quantum evolution of the perturbations. This is a restriction on the possible wave
functionals of the Universe, which should then satisfy condition (\ref{F}). It should
be interesting to investigate situations where entaglement is allowed when the background
is in the quantum regime, which would imply modifications of Eq. (\ref{eqv}) at the bounce. In this
case, condition (\ref{F}) reduces to condition (\ref{Fc}) (no entaglement when the
background becomes classical).

Some future investigations should be to apply the formalism to bouncing models obtained in the framework of quantum cosmology with scalar fields without potential described in Ref.~\cite{colistete} in order to evaluate their spectral index. We will also study the possibility to generalize the simplification of the perturbation equations obtained here to the case of scalar fields with an arbitrary potential term.

\section*{ACKNOWLEDGEMENTS}

We would like to thank CNPq of Brazil for financial support.  We would also like to thank `Pequeno Seminario' of CBPF's Cosmology Group for useful discussions, comments and  suggestions.\\


\end{document}